\documentclass[preprint]
{revtex4}

\usepackage{natbib}
\usepackage{amssymb}
\usepackage{graphicx}

\usepackage{latexsym}

 \usepackage{amsmath, amstext, amsfonts}
 \usepackage[latin1]{inputenc}
 \usepackage{bifchaos}

\begin{document}
\title{A SIMPLE CIRCUIT WITH DYNAMIC LOGIC ARCHITECTURE OF BASIC LOGIC GATES}

\author{I. CAMPOS CANT\'ON}
\affiliation{Facultad de Ciencias, Universidad Aut\'onoma de San Luis Potos\'{\i},
Alvaro Obreg\'on 64, 78000, San Luis Potos\'{\i}, SLP, M\'exico}
\author{E. CAMPOS CANT\'ON} 
\affiliation{Departamento de F\'{\i}sico Matem\'aticas, Universidad Aut\'onoma de San Luis Potos\'{\i},
Alvaro Obreg\'on 64, 78000, San Luis Potos\'{\i}, SLP, M\'exico}
\affiliation{Instituto Potosino de
Investigaci\'on Cient\'{\i}fica y Tecnol\'ogica,\\ Camino a la presa
San Jos\'e 2055, 78216, San Luis Potos\'{\i}, SLP, M\'exico}
\author{J. A. PECINA-S\'ANCHEZ}
\affiliation{Facultad de Ciencias, Universidad Aut\'onoma de San Luis Potos\'{\i},
Alvaro Obreg\'on 64, 78000, San Luis Potos\'{\i}, SLP, M\'exico}
\author{H. C. ROSU}
\affiliation{Instituto Potosino de
Investigaci\'on Cient\'{\i}fica y Tecnol\'ogica,\\ Camino a la presa
San Jos\'e 2055, 78216, San Luis Potos\'{\i}, SLP, M\'exico}

 \begin{abstract}
  \footnotesize
   \noindent
   We report experimental results obtained with a circuit possessing
   dynamic logic architecture based on one of the theoretical schemes proposed by H. Peng and collaborators in 2008.
   The schematic diagram of the electronic circuit and its implementation to
    get different basic logic gates are displayed and discussed. In particular, we show explicitly how to get the electronic NOR, NAND, and XOR gates. The proposed electronic circuit is easy to build because it employs only resistors, operational amplifiers and comparators.\\
    

\noindent Dynamic-logic4.tex \hfill Int. J. Bif. Chaos 20(8) 2547-2551 (2010)
 \end{abstract}

 \maketitle
%


{\bf 1. Introduction}\\

There is great interest in developing new working paradigms in order to
complement and even replace the present statically-wired computer architectures. One of the frontier ideas put forth by Sinha and Ditto in 1998 [Sinha \& Ditto, 1998] is to get
new devices with a dynamic logic architecture that now is called chaos
computing because chaotic (non-linear) elements are employed to get the logic operations [Sinha \& Ditto, 1999; Kuo, 2005; Munakata {\em et al.}, 2002]. 
The main task in this case
is to achieve logic gates (also called cells here) that are able to change their response
according to the control parameter and reconfiguring the device in
order to obtain whatever logic gate. These dynamic logic gates
give us the possibility to build logic chips for the next
generation of computers. Moreover, Sinha and Ditto [Sinha \& Ditto, 1999] 
extended their proposal to encode numbers, perform specific arithmetic operations such as
addition and multiplication, and so on. In a theoretical work of Kuo [Kuo, 2005] 
the logistic map was used as the chaotic element and employed
to emulate logic gates. Currently, there is stimulating research activity in exploiting the logic features of the nonlinear dynamical systems through their electronic implementations [Murali {\em et al.}, 2009; Murali {\em et al.}, 2009; Murali {\em et al.}, 2005; Murali {\em et al.}, 2003]. 

The goal of this work is to present some experimental results that we obtained by means of a simple electronic circuit
with dynamic logic architecture. The circuit was designed based on the first of the three theoretical schemes reported by Peng and collaborators in
order to build dynamic logic gates [Peng {\em et al.}, 2008]. 
The electric diagram of that scheme is easy to implement in our
electronic circuit and the employed components can be acquired in any electronic store.\\




{\bf 2. Electronic Dynamical Logic Gate}\\

The discrete dynamical system used in [Peng {\em et al.}, 2008] 
to simulate dynamic logic gates is the following:

\begin{equation}
  \label{ec01}
  \begin{array}{l}
x(n+1) = C(I_0 + I_1)x(n)\\
y(n+1) = Ky(n)\\

I_{out}=\left\{%
\begin{array}{ll}
    1, & \hbox{if $|y(m) - x(m)| < \beta$} \\
    0, & \hbox{otherwise,} \\
\end{array}%
\right.
\end{array}
 \end{equation}
where $C$ and $\beta$ are positive constants, $K$ is the
bifurcation parameter which acts as the logic-gate controller,
$m$ is the number of iterative steps, $I_0$ and $I_1$ are  the
input logic signals, and $I_{out}$ is the output signal.

The dynamic logic gate given by equation \eqref{ec01} is
controlled through the $K$ parameter according to the intervals
given in table \ref{T_Kvalue} and restricted to $\frac{C}{2} <
\beta \leq C$, $m = 1$, see [Peng {\em et al.}, 2008]. 

\begin{table}
  \centering
\begin{tabular}{|c|c|}
  \hline
  When K $\in$: & The cell is: \\
  \hline
 $[0,C-\beta]$ & NOR \\
 $ (C-\beta,\beta)$ & NAND \\
  $[\beta,2C-\beta]$ & XOR \\
  $(2C-\beta,C+\beta)$ & OR \\
  $[C+\beta,2C]$ & AND \\
  \hline
\end{tabular}
\caption{The ranges of the parameter $K$ for which the system (\ref{ec01}) behaves like the corresponding basic logic gates in the second column.}\label{T_Kvalue}
\end{table}
%

\begin{figure} 
 \centering
 \includegraphics[height=14.3 cm]{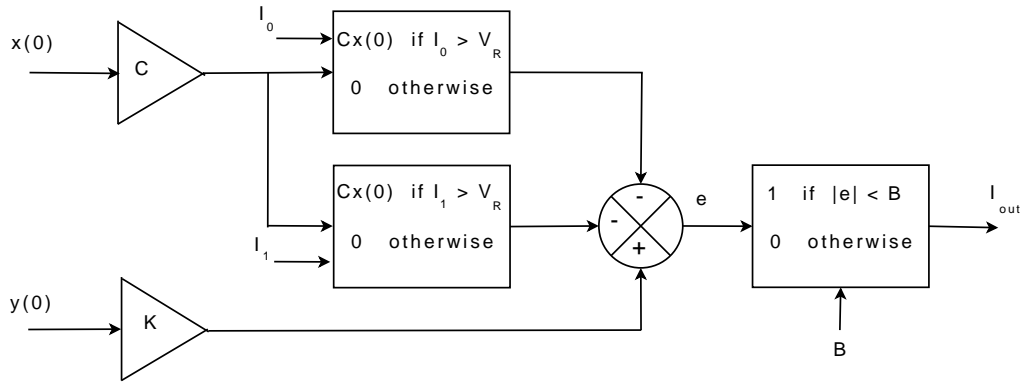}
\caption{\sl Block diagram of the dynamic logic gate.}
 \label{fig_blockDiag}
 \end{figure}

\begin{figure}
\centering
\includegraphics[width= 15 cm, height= 8.0 cm]{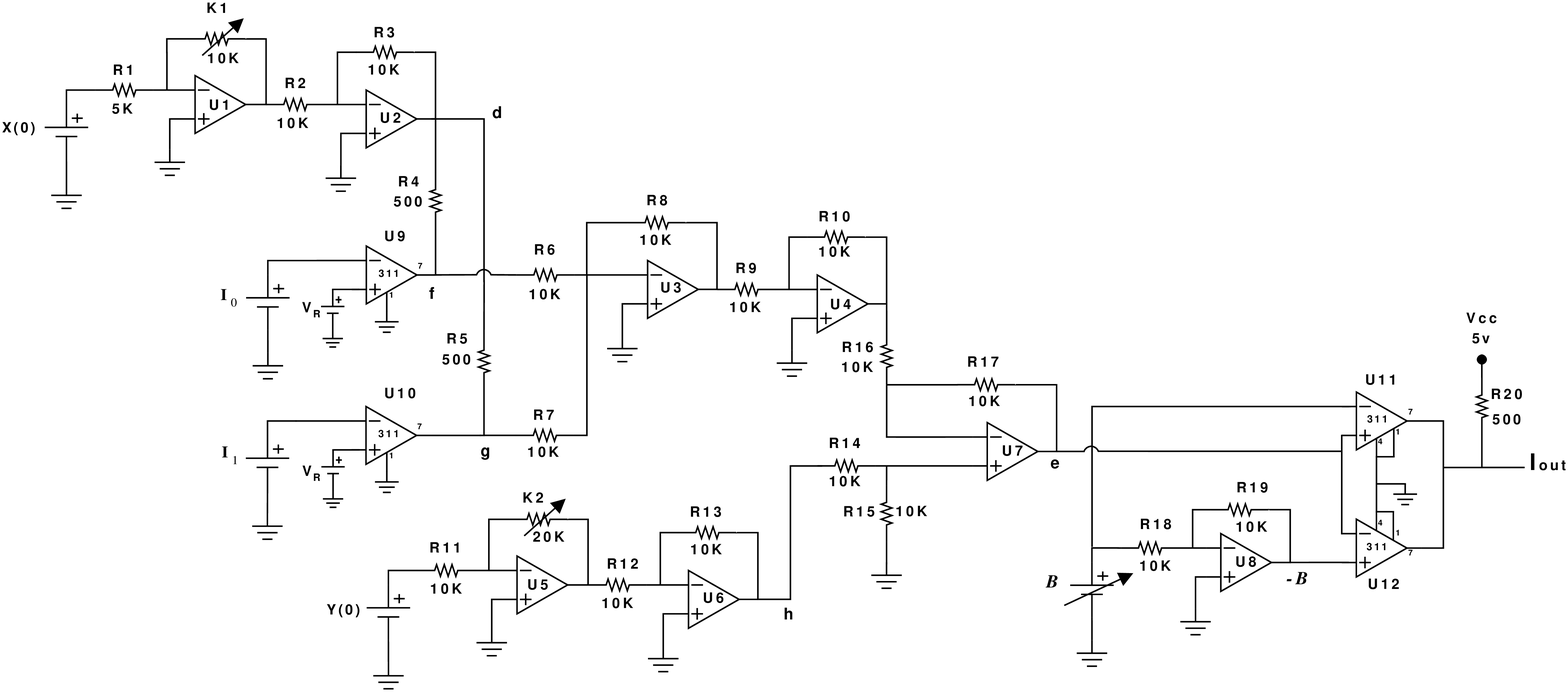}
\caption{\sl Schematic diagram of the dynamic logic gate
electronic circuit.} \label{figcircuito}
\end{figure}

The block diagram of the electronic
circuit of the dynamic logic gate is shown in Fig.~\ref{fig_blockDiag}.
The simplicity of this circuit is due to the fact that the linear mathematical operation of
commutation is performed by the comparators in the switching
blocks, as is shown in Fig.~\ref{fig_blockDiag}. The first
switching block checks if the input $I_0$ is a 0V or 5V and
generates an output equal to 0V or $Cx(0)$, respectively. The
second switching block does the same for the input $I_1$. With
the outputs of the two switching blocks and the input y(0)
the input $e=|y(1)-x(1)|$ of the third switching block is estimated in order to
produce the output $I_{out}$ of the logic gate. The schematic
diagram of the dynamic logic gate circuit is shown in
Fig.~\ref{figcircuito}. It consists of eight operational
amplifiers (from U1 to U8), four comparators (U9 - U12), 21
resistors (from R1 to R21), and two potentiometers $k_1$ and
$k_2$. Assuming ideal performance from all components, the circuit
in Fig.~\ref{figcircuito} is modelled by the following set of
equations in the nodes labelled as d, f, g and h. For the node d,
\begin{equation}
  \label{ec_d}
  d =-\displaystyle\frac{R_3k_1}{R_2R_1}x(0)~.
\end{equation}
The voltages in the nodes f and g are given as follows
\begin{equation}
  \label{ec_fg}
 f =\left\{
   \begin{array}{ll}
     d & \hbox{ if $I_0 > V_R$,} \\
     0 & \hbox{ if $I_0 \leq V_R$,} \\
     \end{array}
     \right.
     \hspace{1.5cm}
 g =\left\{%
\begin{array}{ll}
    d, & \hbox{if $I_1 > V_R$,} \\
    0, & \hbox{if $I_1 \leq V_R$,} \\
\end{array}%
\right.
 \end{equation}
while the voltage at the node h is given by
\begin{equation}
  \label{ec_h}
  h =\displaystyle\frac{k_2R_{14}}{R_{12}R_{13}}y(0)~.
\end{equation}

Combining now the equations for the nodes d, f, g, and h one can get the
voltage in the node e
\begin{equation}
  \label{ec_e}
  e =\left(\displaystyle \frac{1+\displaystyle\frac{R_{18}}{R_{17}}}{1+\displaystyle\frac{R_{15}}{R_{16}}}\right)h
  -\frac{R_{18}R_{11}}{R_{17}R_{10}}\left(\frac{R_9}{R_7}f +\frac{R_9}{R_8}g\right)~.
\end{equation}

Lastly, the outputs of comparators $U_{11}$ and $U_{12}$ as
functions of the voltage in the node e are

\begin{equation}
  \label{ec_Vo}
 U11_0(e) =\left\{%
\begin{array}{ll}
    v_{cc}, & \hbox{if  $\beta > e$,} \\
    0, & \hbox{otherwise.} \\
\end{array}%
\right. \hspace{1cm}
U12_0(e) =\left\{%
\begin{array}{ll}
    v_{cc}, & \hbox{if  $e > -\beta $,} \\
    0, & \hbox{otherwise.} \\
\end{array}%
\right.
 \end{equation}\\

Relating the equations \eqref{ec_d} to \eqref{ec_Vo} and taking the
values for the components given in Table \ref{T_componets} one can get
the output voltage of the circuit. As a
function of $I_0$ and $I_1$, it is given
by the following set of equations:\\

\begin{table} 
  \centering
  \begin{tabular}{|c|c|}
    \hline
    Components & Value \\
    \hline
    Resistors: $R_{1,12}$ & 1 $k\Omega$ \\
    \hline
    Resistors: $R_{2,3,7,8,9,10,11,13,14,15}$ & 10 $k\Omega$\\ 
    \hline
    Resistors: $R_{5,6,21}$ & 500 $\Omega$ \\
    \hline
    Potentiometer: $k_1$ & 10 $k\Omega$ \\
    \hline
    Potentiometer: $k_2$ & 20 $k\Omega$  \\
    \hline
    Op. Amplifiers: $U_{1,\ldots,8}$& TL081 \\
    \hline
    Comparators: $U_{9,\ldots,12}$ & LM311  \\
    \hline
  \end{tabular}
  \caption{The electronic components and their values as employed in the
  construction of the dynamic logic gate electronic circuit.}\label{T_componets}
\end{table}

\begin{equation}
  \label{ecf}
 e =\left\{%
\begin{array}{ll}
    Ky(0)-2Cx(0), & \hbox{if $I_0, I_1 > V_R$,} \\
    Ky(0)-Cx(0), & \hbox{if $I_0 > V_R$ \& $I_1\leq V_R$  or $I_1 > V_R$ \& $I_0\leq V_R$,} \\
    Ky(0), & \hbox{otherwise,} \\
\end{array}%
\right.
 \end{equation}\\
where $K=\displaystyle\frac{k_2}{R_{12}}$ and
$C=\displaystyle\frac{k_1}{R_{1}}$.

\begin{equation}
  \label{ecg}
 I_{out} =\left\{%
\begin{array}{ll}
    v_{cc}, & \hbox{if $|e| < \beta $,} \\
    0, & \hbox{otherwise.} \\
\end{array}%
\right.
 \end{equation}\\



{\bf 3. Experimental Results}\\

The initial conditions $x(0)$ and  $y(0)$ were taken  equal to
$1\,V$. The inputs $I_0$ and $I_1$ of the dynamic logic gate take only
two values $0\,V$ and $5\,V$ in order to represent the binary values 0
and 1, respectively. The voltage $V_R=1\,V$ is the transition
voltage between $0\,V$ (logic 0) and $5\,V$ (logic 1). The different basic
logic gates were obtained varying the value of the potentiometer $k_2$
at fixed value of the potentiometer $k_1$, e.g., for a given
value of the potentiometer $k_1$, tuning the
value of the potentiometer $k_2$ is required in order to get the desirable
logic gate. The value of $\beta$ is adjusted by means of a variable source.
We implemented this design on a printed circuit board (PCB)
manufactured in our laboratory. In the experimental circuit, we
used the TL081 operational amplifiers and the LM311 comparators
supplied with a power source at $\pm 15\,V$ and soldered directly to
the PCB without a socket. The voltage Vdc was supplied by a
variable dc supply source with an output range of 0 - $15\,V$.

We show now how to obtain the different kinds of basic logic
gates. For example, if the potentiometer $k_1$ is equal to the
resistor $R_1$ then $C=1$, and fixing the voltages of $\beta$
to $0.75\,V$, we get $x(0)=1\,V$ and $y(0)=1\,V$. Equations \eqref{ec_e}
and \eqref{ec_Vo} can be rewritten as follows

\begin{equation}
  \label{ec_NR}
 e =\left\{%
\begin{array}{ll}
    K-2, & \hbox{if $I_0, I_1 > 1V$,} \\
    K-1 & \hbox{if $I_0 > 1V$ \& $I_1\leq 1V$  or $I_1 > 1V$ \& $I_0\leq 1V$,} \\
    K, & \hbox{otherwise.} \\
\end{array}%
\right.\\
\end{equation}\
\\
\begin{equation}
  \label{ec_Iout}
 I_{out} =\left\{%
\begin{array}{ll}
    5V, & \hbox{if $|e| < 0.75 $,} \\
    0V, & \hbox{otherwise.} \\
\end{array}%
\right.
 \end{equation}\\

\begin{table} 
  \centering
\begin{tabular}{|c|c|}
  \hline
  When K $\in$: & The circuit behaves \\
  & as the following gate: \\
  \hline
 $[0,0.25]$ & NOR \\
 $ (0.25,0.75)$ & NAND \\
  $[0.75,1.25]$ & XOR \\
  $(1.25,1.75)$ & OR \\
  $[1.75,2]$ & AND \\
  \hline
\end{tabular}
\caption{The intervals of the parameter $K$ and the corresponding basic logic
gate of the circuit .}\label{T_expValues}
\end{table}
The parameter $K$ is controlled by the potentiometer $k_2$ and the
intervals according to the values of the parameters $\beta$ and
$C$ are given in Table \ref{T_expValues}.

We analyze three different cases for the parameter $K$ corresponding to $k_2= 200
\Omega$, $500 \Omega$ and $1 k\Omega$, but other cases can be
investigated in the same way. These values fix the value of $K$
to $0.2$, $0.5$ and $1$, respectively.

\medskip

For $K=0.2$, we have the
following response in the node $e$
\begin{equation}
  \label{ec_k200}
 e_{0.2} =\left\{%
\begin{array}{ll}
    -1.8, & \hbox{if $I_0, I_1 > 1V$,} \\
    -0.8 & \hbox{if $I_0 > 1V$ \& $I_1\leq 1V$  or $I_1 > 1V$ \& $I_0\leq 1V$,} \\
    0.2, & \hbox{otherwise.} \\
\end{array}%
\right.\\
\end{equation}

\begin{figure} 
\centering
  \includegraphics[width=10cm]{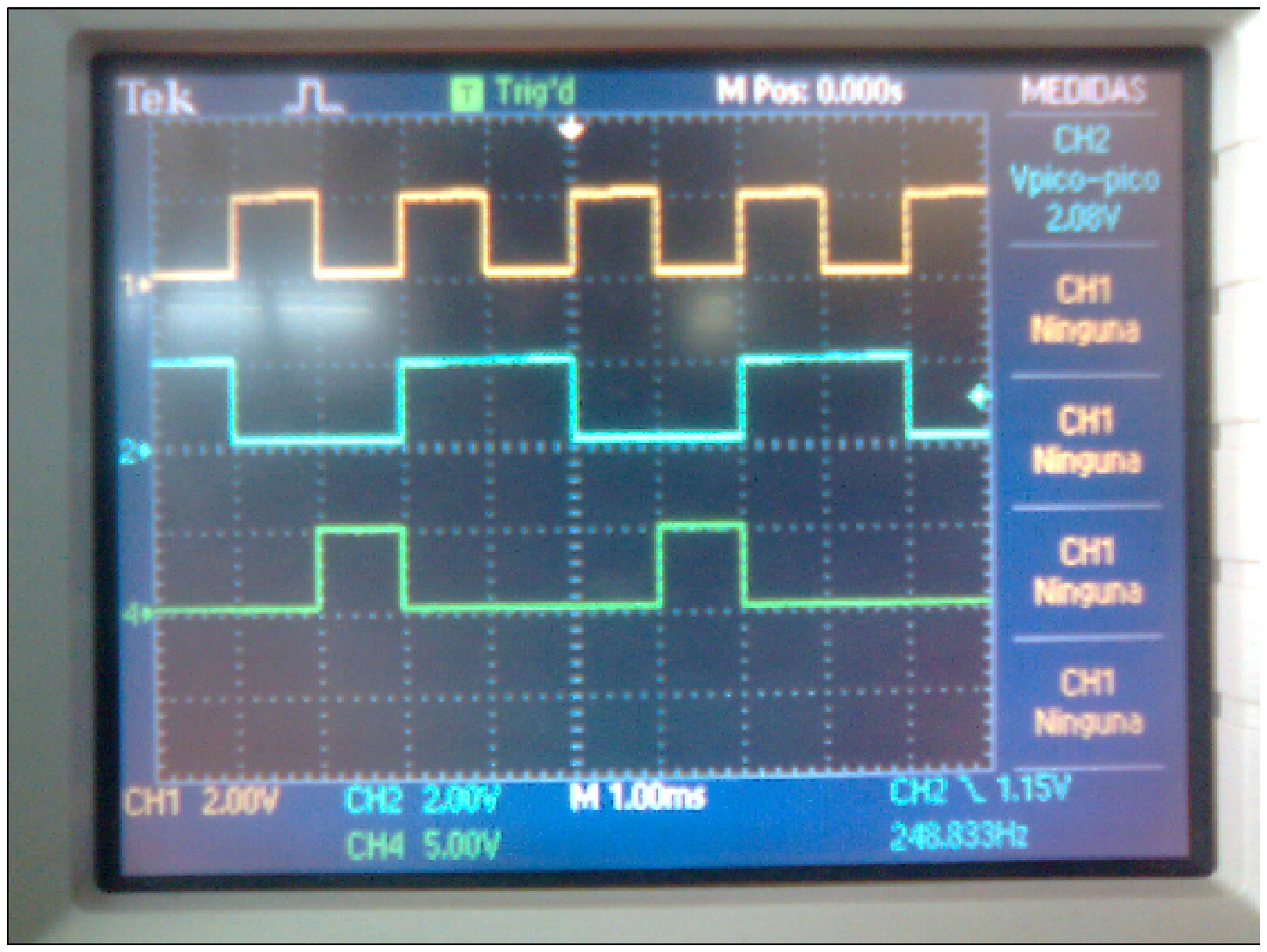}\\
  \caption{The two input signals and the output signal of the NOR gate as seen on the oscilloscope screen.}\label{fig_nor}
\end{figure}

The output given by equation \eqref{ec_Iout} and taking the input
given by equation \eqref{ec_k200} yields the following table

$$
\begin{array}{|c|c|r|c|}
\hline
    I_1(V)&I_0(V) & e(V) & I_{out}(V) \\
    \hline
    0 & 0& 0.2 & 5 \\
    0 & 5& -0.8 & 0 \\
    5 & 0& -0.8 & 0 \\
    5 & 5& -1.8 & 0 \\
    \hline
\end{array}%
$$
Considering that 0V is a logic zero and 5V is a logic one, then
according to the inputs $I_0$, $I_1$ and the output $I_{out}$ the
dynamic logic gate behaves as a NOR gate as can be seen in Fig.~\ref{fig_nor}.

$K=0.5$ corresponds to the interval of the NAND gate, see Table
\ref{T_expValues}. Then, we have
\begin{equation}
  \label{ec_k200}
  \begin{array}{l}
 e_{0.5} =\left\{%
\begin{array}{ll}
    -1.5, & \hbox{if $I_0, I_1 > 1V$,} \\
    -1.5 & \hbox{if $I_0 > 1V$ \& $I_1\leq 1V$  or $I_1 > 1V$ \& $I_0\leq 1V$,} \\
    0.5, & \hbox{otherwise.} \\
\end{array}%
\right.\\
\\

\begin{array}{|c|c|r|c|}
\hline
    I_1(V)&I_0(V) & e(V) & I_{out}(V) \\
    \hline
    0 & 0& 0.5 & 5 \\
    0 & 5& -0.5 & 5 \\
    5 & 0& -0.5 & 5 \\
    5 & 5& -1.5 & 0 \\
    \hline
\end{array}%
 \end{array}
 \end{equation}\\

\begin{figure} 
\centering
  \includegraphics[width=10cm]{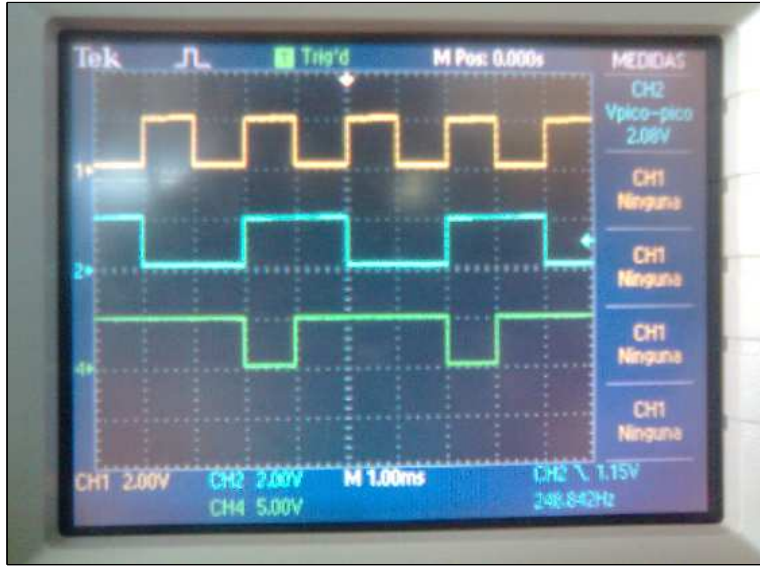}\\
  \caption{The two input signals and the output signal of the NAND gate as seen on the oscilloscope screen.}\label{fig_nand}
\end{figure}

For this case, the experimental input-output signals of the NAND gate are shown in Fig.~\ref{fig_nand}.

\medskip

 For $K=1$,  we have the dynamic logic gate behaving as a XOR gate. The input
 and output of the circuit are shown in the next equation and table
\begin{equation}
  \label{ec_k21000}
  \begin{array}{l}
 e_{1} =\left\{%
\begin{array}{ll}
    -1, & \hbox{if $I_0, I_1 > 1V$,} \\
    0 & \hbox{if $I_0 > 1V$ \& $I_1\leq 1V$  or $I_1 > 1V$ \& $I_0\leq 1V$,} \\
    1, & \hbox{otherwise.} \\
\end{array}%
\right.\\
\\

\begin{array}{|c|c|r|c|}
\hline
    I_1(V)&I_0(V) & e(V) & I_{out}(V) \\
    \hline
    0 & 0& 0 & 0 \\
    0 & 5& 0 & 5 \\
    5 & 0& 0 & 5 \\
    5 & 5& -1 & 0 \\
    \hline
\end{array}%
 \end{array}
 \end{equation}\\

\begin{figure}  
\centering
  \includegraphics[width=10cm]{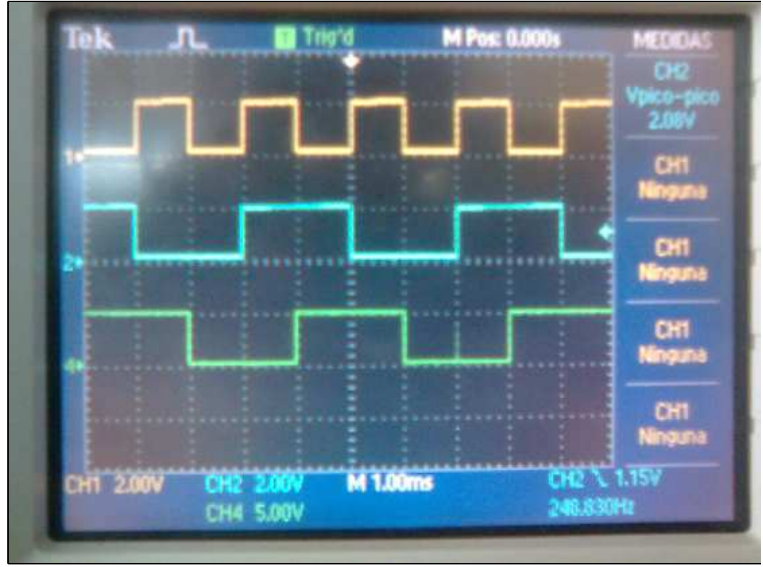}\\
  \caption{The two input signals and the output signal of the XOR gate as seen on the oscilloscope screen.}\label{fig_xor}
\end{figure}

 The experimental results are presented in Fig.~\ref{fig_xor}.

The other logic gates can be checked in the same way for the
different values of the parameter $K$.\\

\newpage



{\bf 4. Conclusions}\\

A very simple dynamic logic gate electronic circuit has been
described here together with its implementation using only analog
components such as operational amplifiers, comparators and
resistors. Its experimental behavior was tested and compared with
the numerical behavior given by the
mathematical model of the dynamical logic gate that has been taken from the work of Peng and collaborators [Peng {\em et al.}, 2008]. 
Choosing different values of the potentiometer $k_2$ offers us the possibility to get different
logic gates: NOR, NAND, XOR, and others. Such circuit realizations have many
potential applications in chaos computing. Finally, we notice that
our design can be manufactured in just one chip because the final
electronic circuit contains only semiconductors and passive components.

\bigskip
\medskip


\noindent {\bf References}
%

\noindent Sinha, S., Ditto, W.L. [1998] ``Dynamics based computation," {\em Phys.
Rev. Lett.} {\bf 81}, 2156-2159.

\noindent Sinha, S., Ditto, W.L. [1999] ``Computing with distributed chaos," {\em Phys.
Rev. E} {\bf 60}, 363-377.

\noindent Kuo, D. [2005] ``Chaos and its computing paradigm," {\em IEEE Potentials} {\bf 24}, 13-15.

\noindent Munakata, T., Sinha, S., Ditto, W.L. [2002] ``Chaos computing:
implementation of fundamental logical gates by chaotic elements,"
{\em IEEE Trans. Circuits Syst., I: Fundam. Theory Appl.} {\bf 49}, 1629-1633.

\noindent Murali, K., Sinha, S., Ditto, W.L., Bulsara, A.R. [2009]
``Reliable logic circuit elements that exploit nonlinearity in the presence of a noise floor,"
{\em Phys. Rev. Lett.} {\bf 102}, 104101.

\noindent Murali, K., Miliotis, A., Ditto, W.L., Sinha, S. [2009]
``Logic from nonlinear dynamical evolution,"
{\em Phys. Lett. A} {\bf 373}, 1346-1351.

\noindent Murali, K., Sinha, S., Mohamed, I.R. [2005]
``Chaos computing: experimental realization of NOR gate using a simple chaotic circuit,"
{\em Phys. Lett. A} {\bf 339}, 39-44.

\noindent Murali, K., Sinha, S., Ditto, W.L. [2003]
``Implementation of NOR gate by a chaotic Chua's circuit,"
{\em Int. J. Bif. and Chaos} {\bf 13}, 2669-2672.

\noindent Peng, H., Yang, Y., Li, L., Luo, H. [2008] ``Harnessing piecewise-linear
systems to construct dynamic logic architecture," {\em Chaos} {\bf 18},
033101.



 \end{document}